\documentclass[preprint,12pt]{elsarticle}

\usepackage{color}
\usepackage{graphicx}
\usepackage{amssymb}
\journal{Physics of the Dark Universe}

\begin{document}

\begin{frontmatter}

\title{Planck and the local Universe: quantifying the tension}
\author[1,2]{Licia Verde}
\author[3,4]{Pavlos Protopapas}
\author[1,2]{Raul Jimenez}
\address[1]{ICREA \& ICC, University of Barcelona (UB-IEEC), Marti i Franques 1, 08034, Barcelona, Spain}
\address[2]{Theory Group, Physics Department, CERN, CH-1211, Geneva 23, Switzerland}
\address[3]{Harvard Smithsonian Center for Astrophysics, Cambridge, MA 02138, USA}
\address[4]{Institute for Applied Computational Science, School of Engineering and Applied Sciences, Harvard University, Cambridge, MA 02138, USA}
\begin{abstract}
 We use the latest {\it Planck} constraints, and in particular constraints on the  derived parameters (Hubble constant and age of the Universe) for the local universe  and  compare them with local measurements of the same quantities. We propose a way to quantify whether cosmological parameters constraints from two different experiments are in tension or not. Our statistic, $\mathcal{T}$, is an evidence ratio and therefore can be interpreted with the widely used Jeffrey's scale.  We find that in the framework of the $\Lambda$CDM model, the {\it Planck} inferred  two dimensional, joint,  posterior distribution for  the Hubble constant and age of the Universe is in ``{\it strong}'' tension with the local measurements;  the odds being $\sim 1\!\!:\!\!50$.
We explore several possibilities for explaining this tension and examine the consequences both in terms of unknown errors and deviations from the $\Lambda$CDM model. In some one-parameter $\Lambda$CDM model extensions,  tension is reduced whereas in other extensions, tension is instead increased. In particular, small total neutrino masses are favored and  a total neutrino mass above  0.15 eV makes the tension ``{\it highly significant}'' (odds $\sim 1\!\!:\!\!150$). A consequence of accepting this interpretation of the tension is that  the degenerate neutrino hierarchy is highly disfavoured by cosmological data and  the  direct hierarchy is slightly favored over the inverse.
\end{abstract}

\end{frontmatter}
\section{Introduction}
\label{sec:intro}

Cosmic Microwave Background  (CMB) data have been crucial to define and confirm the currently  favored cosmological model: a  flat cosmological constant--dominated, cold dark matter model, $\Lambda$CDM. It is important to keep in mind that CMB observations predominantly probe the physics of the early Universe ($z \gtrsim 1100$). When these observations are interpreted in terms of  the standard cosmological parameters, defined at $z = 0$, an extrapolation is needed, which is done within a given cosmological  model.  
In our previous work \cite{local}, we argued that local, model-independent measurements of cosmologically-relevant quantities can be used to test the self-consistency of the currently favored cosmological model and to constrain deviations from it. Direct measurements of the Hubble constant and age of the Universe  are especially suited to this aim.
In fact, these measurements  have a long history and have now  reached a  level of precision and accuracy that make them competitive with other cosmological observations. 
If local and high-redshift measurements are to be combined to constrain cosmological parameters within a given model, the well established framework of Bayesian parameter inference can be used. However, if the two sets of measurements are to  be used to distinguish between models then  Bayesian model selection  and Bayesian Evidence\footnote{Here we use capital "Evidence" for the Bayesian quantity to distinguish it from the colloquial {\it evidence}.}  (model-averaged likelihood) should  be used \cite{Cox}. Bayesian methods in cosmology have  been used for almost two decades now (see \cite{Jaffe96,Hobson,Liddle,Parkinson,Trotta1,Trotta2}  and references therein). 

Ref.~\cite{local} used pre-{\it Planck}  state of the art data, so it is a natural extension to that work to consider the  post-{\it Planck} \cite{Planckbasicresults} state-of-the-art cosmological data.
The {\it Planck} team official analysis pointed out that local direct measurements of the Hubble constant seem to be at odds with CMB  data when interpreted within the context of the  $\Lambda$CDM model. Post-{\it Planck} CMB data are however not at odds with other cosmological measurements (e.g., Baryon Acoustic Oscillations). 
This conclusion  arises from a parameter-estimation analysis, where the best fit  value of the  Hubble constant  extrapolated  from CMB data is $\sim 2.5 \sigma$ away from the direct measurement.

We will examine these findings in the framework of  Bayesian model  selection and  discuss their implications for cosmology. We will also consider measurements of the age of the Universe and investigate whether, from a model selection point of view, the tension between local and high redshift measurements  disfavors the $\Lambda$CDM model in favor of a more complex model. 

This paper is organised as follows:  In Sec.~\ref{sec:data}, we present the data sets and combinations of data sets we will use. In Sec.~\ref{sec:methods},  we review different statistics that measure the ``distance" or the ``difference" between two distributions. We present our statistic of choice that stems from the Bayesian evidence ratio and  is suited to assess whether two posterior distributions from two different experiments are in tension or not. 
We present our results in Sec.~\ref{sec:results}, where we examine the tension between local and high redshift cosmological  measurements.  We then  explore several possibilities for this tension and examine the consequences both in terms of unknown errors and deviations from the $\Lambda$CDM model.  Finally, we conclude in Sec.~\ref{sec:conclusions}. In  \ref{sec:kld}, we report the  Kullback-Leibler divergence  between Planck and WMAP for selected models and parameters.

\section{Data}
\label{sec:data}
We consider the {\it Planck} mission CMB data with local measurements of the Hubble parameter $H_0$ and of the age of the Universe $t_U$.
\subsection{CMB data}
The Planck collaboration, along with the nominal mission  temperature data, has also released  the outputs of the Markov Chains Monte Carlo (MCMC) used to sample from the space of possible cosmological parameters. These MCMCs have been used by the team to generate estimates of the posterior mean of  cosmological parameters, along with their confidence intervals \cite{Planckparameterspaper}. Here, we use the publicly available outputs of the {\it Planck}  team MCMCs. 
The {\it Planck}  data  have been analyzed by the team in several ways:  using only  Planck temperature data on scales corresponding to multipoles $\ell <2500$ ({\it Planck}), using  Planck data in conjunction with WMAP polarization data at low $\ell$ ({\it WP}), and using Planck data, WMAP data and also include the measurement of the lensing potential, which was reconstructed from the Planck temperature maps themselves through the measurement of the four-point function ({\it lensing}). 
In some cases, data from high-resolution but partial sky ground-based CMB experiments (ACT and SPT) have also been included ({\it highL}).  In combination with Planck  these data better constrain the foreground-model parameters. 
Here however, we concentrate on the {\it Planck +WP} data combination.  Except in the Appendix, we will not consider  the {\it lensing} information in line with the considerations presented in \cite{Feeney2}; the combination  {\it Planck +WP+highL}  predicts a value for the lensing amplitude which is about $2 \sigma$ higher than the value measured from the convergence power spectrum \cite{Plancklensingpaper,Planckparameterspaper} and  the origin of this tension is not yet fully understood.

\subsection{Local measures: $H_0$ and $t_U$}
For the local data we follow exactly Ref.~\cite{local}.  We combine the values reported in  Ref.~\cite{Riess/etal:2011} and \cite{Freedman/etal:2012} in  a ``world average" where the  central value is  given by the variance-weighted mean and the  error conservatively  given by the average of the errors: $H_0=74.08 \pm 2.25$ Km s$^{-1}$Mpc$^{-1}$.

Estimates of the age of the Universe can be obtained from the ages of the oldest objects and in particular from the  ages of the oldest stars since these objects form very shortly after the Big Bang.
Accurate dating of globular clusters  has been the subject of active investigation for decades but the error bars were, for a long time, large. It is only relatively recently that error-bars have become smaller due to better estimates of their distances \cite{chabkrauss} or the use of distance independent methods \cite{Jimenez:1996at}.
Recently, it has become possible to use single stars to estimate $t_U$; accurate distances using direct parallax measurements were obtained for nearby sub-giant stars.  In particular the star HD 140283 is a sub-giant moving off the main sequence, so  its luminosity is a very good  age-indicator.

Consequently, in order to constrain the current age of the universe, we use recent determinations of the ages of the oldest stars in the Milky Way. We use two kind of measurements: the age of the nearby sub-giant HD-140283 and the ages of the oldest globular clusters. The age of HD-140283 has been accurately measured by \cite{Bond} using HST parallaxes and spectroscopic determinations of its chemical abundance. In addition, they have used state of the art stellar  evolutionary models and carried out a careful and  extensive  error budget. The age of HD-140283 is determined to be $14.5 \pm 0.8$ Gyr (including systematic errors, which dominate the error bar). Additionally, the ages for some of the most metal poor Milky Way globular clusters (NGC 6397, NGC 6752, and 47 Tuc) have been determined by Ref.~\cite{Gratton03}. Taking into account the revised nuclear reaction rate for  $^{14} N (p, \gamma)^{15} O$ of the CNO burning cycle \cite{Imbriani}, we obtain  an age $14.2 \pm 0.6 (\pm  0.8 \,{\rm systematics})$ Gyr. In what follows we will linearly add   these (random and statistics) two sources of errors. 

It is remarkable that age determinations for such different systems are in such a good agreement. There are three main ingredients that dominate the error budget of stellar ages: distance, chemical composition and theory of stellar evolution. The last one has been studied with extreme care in the last decade and major  improvements have been made. It is very unlikely that stellar evolution theory needs any further  significant revisions
; the contribution to the age error budget from stellar evolution theory is now negligible contributing  to about 1\% to the total age uncertainty. The two dominant error sources remain distance and chemical composition. 

The distance uncertainty can be efficiently  removed by obtaining trigonometric parallaxes to the oldest stars or globular clusters. This has been the case for HD-140283. However, for the globular clusters it is not yet possible to obtain trigonometric parallaxes and distances have to be obtained by indirect methods. 
Currently the uncertainty budget in the age determination of galactic globular clusters is dominated by its distance estimation, which is done via indirect methods (sub-dwarf fitting).  

On the other hand, the dominant source of error of nearby sub-giants  is the chemical composition of the stars. Current 10-meter class  telescope observations provide abundance with an accuracy of $0.1$ dex, which translates into a $\sim 5$\% uncertainty when estimating the age of the oldest stars. 

The ages determinations of nearby sub-giants and globular clusters are dominated by different  and independent systematics (chemical abundance and distance respectively) and  we therefore combine the two above measurements by inverse variance weighting.
Recall that the ages of the oldest objects at $z=0$ only provide a lower estimate of $t_U$. However, we know that the first generation of stars formed at $z > 20$ \cite{Zoltan}. At this redshift the age of the universe is only $\sim 0.1-0.2$ Gyr, which is much smaller than the current age estimation errors. 
This formation time-lag  will only play a role when age estimations uncertainties are reduced by at least a factor of five, to the $0.7$\% level. In our analysis we convert the age of the star to the age of the Universe by adding to the stellar age $0.15 \pm 0.05$ Gyr, assuming a Gaussian distribution (cutting negative tails when appropriate).  We obtain the following estimate for the age of the Universe:  
$t_U=14.4 \pm 0.7$. Here we assume Gaussian distributions for both $H_0$ and $t_U$ measurements.

It is important to note that the  error-budget in the age determination  will change dramatically with the launch of the GAIA satellite in October 2013. GAIA will obtain trigonometric parallaxes to globular clusters and identify a large number of metal-poor sub-giants. Therefore, the only dominant source of error remaining will be the chemical composition of the stars. A way to improve on this will be to obtain longer integrations or use different line diagnostic for the  metallicity in the infrared. Thus the dominant uncertainty in the error budget for the age of the local universe, will be the chemical composition reaching an error-floor of $\sim 2-3$\%; see \cite{local} for discussion.   
  
\section{Methods}
\label{sec:methods}

The questions we wish  to address are the following.  {\it a)} We have two measurements of cosmologically interesting quantities in the form of a --two or higher dimensional-- posterior  distribution. In our application these two measurements are  the  high-redshift, model-dependent  joint distribution of $H_0$ and $t_U$, and the local, cosmological-model-independent one.  In the Bayesian framework, how would one quantify whether these two measurements are or not in agreement (tension)? In other words, if the null hypothesis is that the two measurements  are ``sampled" from the base model  adopted, when should the null hypothesis be  abandoned? {\it b)} If the answer to {\it a)} is that the two measurements are in tension, then Bayesian model selection can be used to study extensions to the base model adopted and  select which is the favored model. Alternatively the detected tension might indicate the presence of  unaccounted for, residual systematic errors. Possible options at this point are: discredit  the measurement most likely affected by systematics or artificially increase its errors.  If instead no tension is detected the measurements can be combined to perform, for example, joint parameter estimation. 

 Clearly to address this type of question a measure of distance or difference between   two distributions should be used. There are many  statistics that quantify the difference between two distributions. The most widely used are the Kullback-Leibler divergence \cite{KL}, the Jeffrey's divergence \cite{Jeffreys}, and  Jensen-Shannon divergence \cite{Lin}. These are well rooted in  information theory and they measure the difference between two probability distributions, say ${\cal P}$ and ${\cal Q}$, and therefore are suited to quantify how well ${\cal P}$ approximates ${\cal Q}$ or the information content that ${\cal Q}$ adds to ${\cal P}$. In fact the Kullback-Leibler divergence is not symmetric (clearly,  the information content that ${\cal Q}$ adds to ${\cal P}$ is not the information content that ${\cal P}$ adds to ${\cal Q}$) and the Jeffrey's divergence and Jensen-Shannon divergence are two approaches to symmetrize it. We will explore the application to cosmology of the entropy-based  Kullback-Leibler divergence measure in the Appendix.  For our question {\it a)} at hand we know {\it a priori} that  we will be comparing two different distributions (in fact we are comparing different experiments). We want to know wether the best fit  values are consistent given the shape of the respective distributions, in other words  we need the Bayesian (multi-dimensional) parallel of the standard equivalency test \footnote{Say that we have two measurements (A and B), with errors ($\Delta A$ and $\Delta B$) of the same quantity,  the standard equivalency test says that  $A$ and $B$ are consistent within the errors if $|A-B|\le \sqrt{\Delta A^2+\Delta B^2}$.}. We argue that this is given by the Evidence ratio as follows.
 
 \subsection{Evidence for tension}
 \label{sec:evidencetension}
 Imagine we have performed two experiments: $A,B$ and for each experiment
we produce a posterior $P_{A,B}(\theta|D_{A,B})$ where $\theta$ represents the parameters
of the model and $D_{A,B}$ represents the data from experiments $A,B$ respectively.  Let us also assume that for producing both posteriors we have used the same, uniform priors over the same ``support", x, i.e., $\pi_A=\pi_B=\pi$, $\pi=1$ or $0$ and  therefore $\pi_A \pi_B=\pi$.

Let $H_1$ be the (null) hypothesis that both experiments measure the same quantity, 
the models are correct and there are no unaccountable errors.  In this case, the two experiments will produce two posteriors,  which, although can have different (co)variances, and different distributions, have means that  are in agreement.
The alternative hypothesis, $H_{\neg 1}$ is when the two experiments, for some unknown reason,   do not agree, either because of systematic errors or because  they are effectively measuring different things or the model (parameterization) is incorrect. In this case, the two experiments
will produce two posteriors with two  different means and  different
variances.

To distinguish the two hypothesis we use the  Bayes factor,  that is  the ratio of the Evidences.

 In any practical application, the absolute normalization of the posteriors is often unknown, but we can still  work as follows. We define:
 \begin{equation}
\int P_A P_B dx =\lambda \int {\cal L}_A{\cal L}_B \pi_A\pi_Bdx =\lambda\int {\cal L}_A{\cal L}_B \pi dx =\lambda E={\cal E}\,,
\end{equation}
where ${\cal L}$ denotes the likelihood and $\lambda^{-1}=\int{\cal L}_A\pi_Adx \int{\cal L}_B\pi_Bdx'$.  $E$ is the Bayesian Evidence for the joint distribution, thus ${\cal E}$ is akin to an unnormalized Evidence.
Operationally, to deal with the (difficult to compute) normalization factor  let us now imagine that we can perform a  {\it translation} (shift) of  (one or both of) the distributions in $x$ and let us define $\bar{P}_A$ the shifted distribution. This translation changes the location of the maximum but does not change the shape or the width of the distribution. 
If the maxima of the two distributions coincide then 
\begin{equation}
\int \bar{P}_A \bar{P}_B dx=\bar{\cal E}|_{{\rm max}A={\rm max B}}\,.
\end{equation}
This can be considered our ``straw man" null hypothesis. 
 As the distance between the maxima increases (but the shape of the distributions remains the same),
\begin{equation}
\int \bar{P}_A \bar{P}_B dx=e<\bar{\cal E}\,,
\end{equation}
and eventually $e\longrightarrow 0$ as the two distributions diverge. Clearly the Evidence ratio for the (null) hypothesis $E_1$ is ${\cal E}/\bar{\cal E}|_{{\rm max}A={\rm max B}}$, as the normalization factors $\lambda$ cancel out, and  the Evidence ratio for the alternative $H_{\neg 1}$ is its reciprocal.
We therefore introduce:
\begin{equation}
\mathcal{T}=\frac{\bar{\cal E}|_{{\rm max}A={\rm max B}}}{\cal E} \, ,
\label{eq:tension}
\end{equation}
  which denotes the degree of tension that can be interpreted in the  widely used (slightly modified, \cite{kassraftery}) Jeffrey's \cite{Jeffreys} scale (Tab. \ref{tab:scale}). $\mathcal{T}$ indicates the odds:   $1\,:\,\mathcal{T}$  are the chances  for the null hypothesis. In other words, a large tension mens that the null hypothesis (${\rm max} A$ = ${\rm max} B$)  is  unlikely.
  \begin{table}[tb]
\centering
 \begin{tabular}{ccc}
 \hline
$  \ln \mathcal{T} $ & interpretation & betting odds\\
 \hline
 $<1$ & not worth a bare mention, not significant &$<3:1$ \\
 $1-2.5$ & substantial & $\sim 3:1$\\
 $2.5-5$ & strong &$>12:1$ \\
 $>5$ & highly significant & $>150:1$\\
  \hline
\end{tabular}
 \caption{The slightly modified Jeffreys' scale we use for interpreting the tension $\mathcal{T}$.}
 \label{tab:scale}
\end{table}

In this scale, $\ln \mathcal{T}<1$ is {\it not significant},  if $1<\ln \mathcal{T}<2.5$ the evidence is {\it substantial}, becomes {\it strong} only if $2.5<\ln \mathcal{T}<5$ and {\it highly significant} if $\ln \mathcal{T}>5$.  

This scale is empirically calibrated,  and should be used only as a guide as it  introduces sharp decision-making boundaries. Here we use the boundaries as a  loose classification of the degree of  tension; we also use Jeffreys' nomenclature.

To give an intuition about the meaning of $\ln \mathcal{T}$ values, consider  two Gaussian  distributions with unit variance: a shift of the central value of one of the two distributions  of $2 \sigma$ would give $\ln \mathcal{T}=1$ which is the threshold between {\it not significant} and {\it substantial}; a shift of more than  $3 \sigma$ would yield `{\it strong} tension and  of  more than $>4.5 \sigma$ to give {\it highly significant} tension.

In some practical applications the shift needed to compute the numerator of $\mathcal{T}$ may be  slightly incorrect or  at least misleading. For example, we know that for CMB data the covariance matrix depends on the assumed cosmology because of cosmic variance. Therefore  a rigid translation of the distribution is strictly incorrect. One could in principle imagine an extreme case where shifting $P_A$ so that the maximum coincides with that of $P_B$ gives a highly significant Evidence but instead shifting $P_B$ does not.  

However, for the practical applications we can think of, in the era of precision cosmology,  this effect is small, or if it is large it means that the shift is large  and the  Evidence for tension will be  highly significant anyway. Therefore this effect  will not drastically change the  interpretation  of the $\mathcal{T}$ value.  Nevertheless, in what follows, we  shift the local measurements distribution, which is not affected by cosmic variance and thus does not depend (too strongly) on cosmology. 

When computing $\mathcal{T}$ in a practical application, there is a delicate point to bear in mind:  the above relies on having a uniform prior on $H_0$ and $t_U$.   Here we wish to use the output of MCMCs which were performed with uniform priors on other parameters (not $H_0$ and $t_U$). The relation between these parameters and $H_0$,  $t_U$ is non-linear, thus in the MCMCs the prior in not uniform in $H_0$,  $t_U$. As a consequence, the MCMC outputs in principle, cannot simply be importance-sampled  and then   used to perform the above integrals by Monte Carlo integration i.e.,  simply adding up the MCMC (updated) weights. Of course, a change of variables can be made by re-weighting the chains outputs by the Jacobian of the transformation. This effect for the data sets considered here is, however, small.  Alternatively the posterior surface in the $H_0$,  $t_U$  sampled by the MCMC can be fit by a smooth surface. This surface provides then a functional form for the posterior which can be integrated to compute ${\cal E}$ and  $\mathcal{T}$. We use the latter approach.

With this type of analysis we are entering the regime of ``meta analysis" which has  an extensive literature mostly in the medical field.

Logically, if ${\cal T}$ is small and  the tension is not significant the  results of the experiments can be combined and a joint analysis can be safely  performed. However if ${\cal T}$ is large (e.g., {\it strong} or {\it highly significant}), then it is an indication that either {\it a)} one of the two experiments is affected by errors (systematic or statistical) that are unaccounted for or that  {\it b)} the underlying model used is incorrect and must be extended.

The situation described in case {\it a)} has been considered before in the cosmology literature \cite{Bridle, Hobson}.  The authors advocate introducing ``meta parameters"  describing possible systematic shifts or statistical errors and marginalize over them.  The detailed implementation in the Bayesian context is numerically very heavy, but  the upshot is that  in practice such an approach leads to  down-weighting the  discrepant measurements (i.e., effectively increasing the corresponding error-bars) but still combining them.  
A simplified approach (which is the one we will pursue here) is to increase the error-bars of the  discrepant measurement(s) until the tension ${\cal T}$ is not significant. After that we can  analyze   jointly the data sets. We will refer to this case as ``blame the measurements".

The situation described in {\it b)} is similar  to what is being done extensively in cosmology when implementing model selection in the Bayesian framework. There is a simpler model (typically the 6 parameters, flat,  $\Lambda$CDM--``base''-- model) and simple extensions of it, where one or two quantities, which have fixed values in the ``base" model, are promoted to parameters of the model. In this case model selection is carried out by computing the Bayes factor, or evidence ratio, between the two models. The question this approach addresses is the model selection question:  ``is  the introduction of  the extra parameter(s) warranted by data?".  However, there is also another question we can ask (and we are interested in addressing here):  ``Does the introduction of the extra parameter reduces the tension as defined in Eq.~\ref{eq:tension}?"
And further, ``what are the  fixed values (if any) of the extra parameter that would make the tension not significant?".
In this paper we will concentrate on the last two questions. We will refer to this case as  ``blame the model".

\section{Results}
\label{sec:results}
We begin by repeating  some of the key steps of the analysis of Ref.~\cite{local} using the  updated state-of-the-art data.  
\begin{figure}[h]
        \centering
                \includegraphics[scale=0.5]{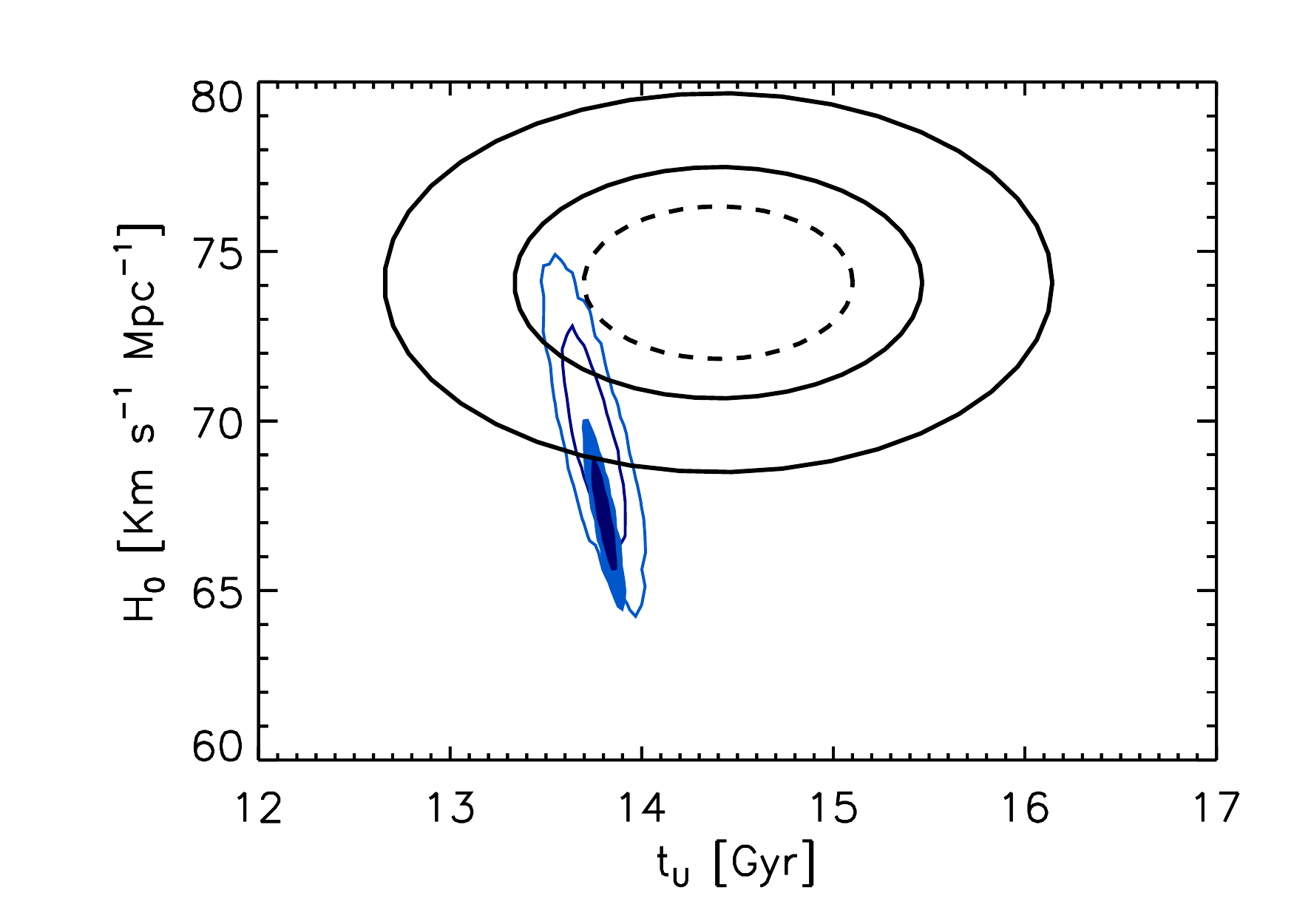}
   \caption{\label{fig:LCDMwmapplanck} Constraints (1 and 2 $\sigma$ joint) in the  $t_U$--$H_0$ plane from local measurements (black solid contours, the dashed contours corresponds to the  single parameter, marginalized constraint) and CMB data (blue). The transparent set of contours correspond to WMAP and the filled contours to Planck. 
}
\end{figure}
Fig.~\ref{fig:LCDMwmapplanck} shows constraints on the $t_U$--$H_0$ plane from local measurements and from CMB in the framework of the standard $\Lambda$CDM model; both WMAP and {\it Planck} constraints are shown. In \ref{sec:kld} we quantify how much information Planck has added to WMAP for this particular parameter combination within the  $\Lambda$CDM model. For now, we can appreciate that the {\it Planck} central value has shifted compared to WMAP's. This shift is well within the WMAP $1$ $\sigma$ confidence region, but the reduced {\it Planck} error-bars mean that now the $1$ $\sigma$ confidence regions of CMB and local measures do not overlap (only the $2$ $\sigma$ joint  still do). This  represents  the above mentioned ``tension". We will return on this in Sec.~\ref{sec:tension} below.

\begin{figure}[h]
        \centering
                \includegraphics[scale=0.53]{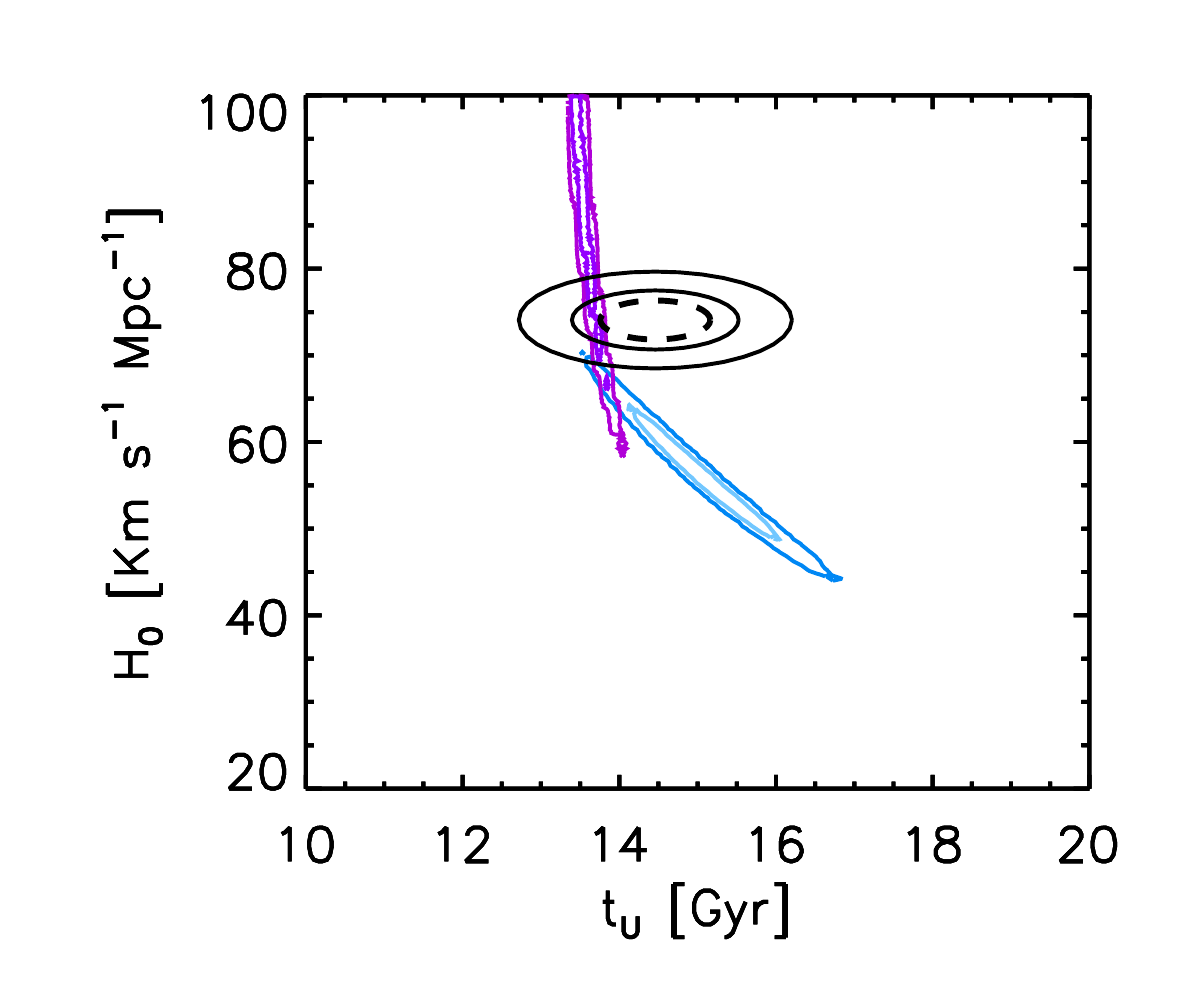}
                \includegraphics[scale=0.53]{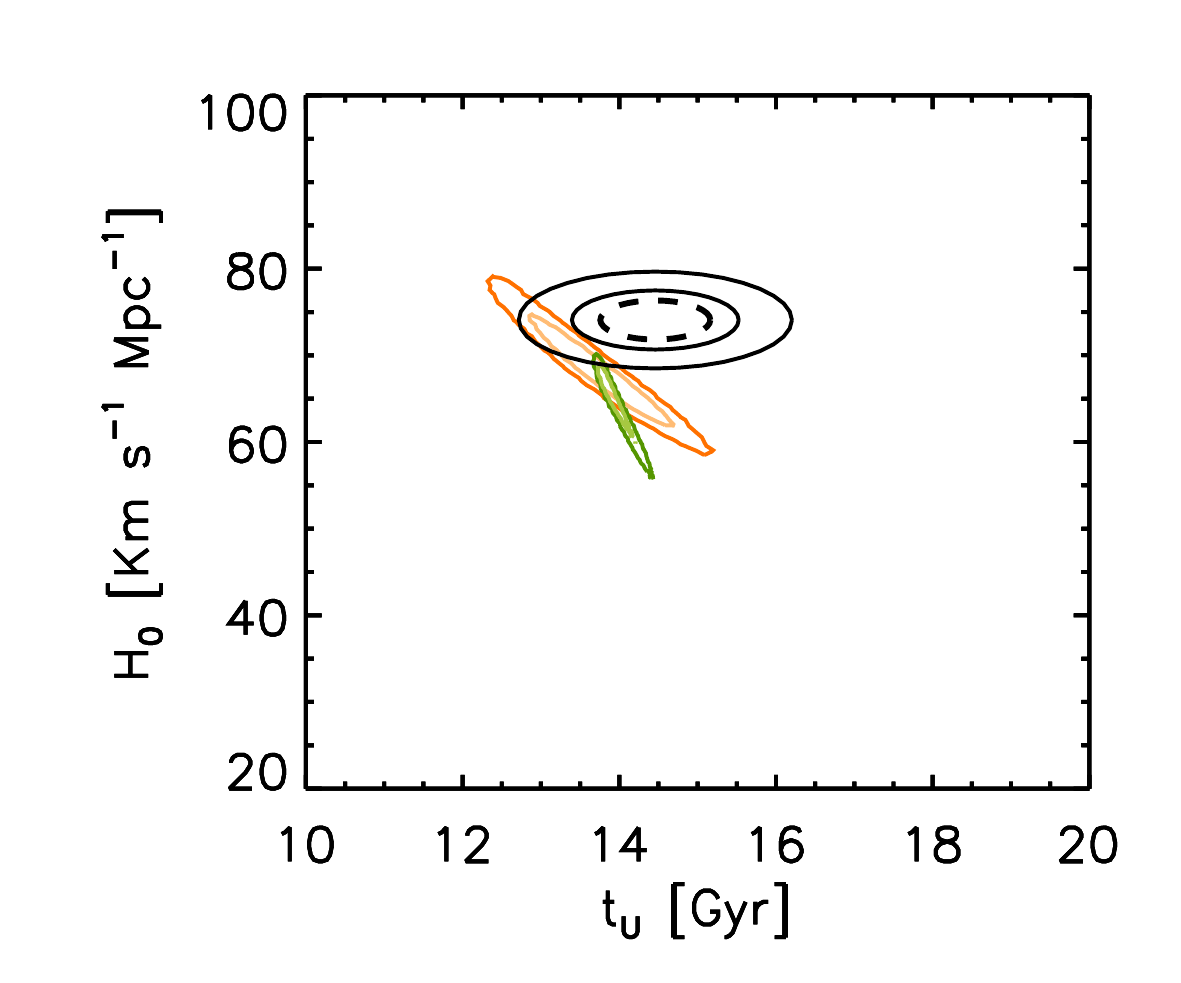}
               \caption{\label{fig:all} Left panel: blue: curvature extension to the $\Lambda$CDM model,  magenta: equation of state parameter for the dark energy $w$ extension to the $\Lambda$CDM model. Right panel: non standard neutrino properties. Green: neutrino mass and primordial helium content extension and Orange: number of effective neutrino species and primordial helium content extension. The plot range and color scheme have been chosen so these figures  can be compared  directly, at a glance, with Fig. 3 of Ref.~\cite{local} for a direct comparison with WMAP.}
\end{figure}

The smallness of the {\it Planck} allowed region on this plane is due to the assumption of the $\Lambda$CDM model. In Fig.~\ref{fig:all} we show how this changes for simple (one or two parameters) extensions to the   $\Lambda$CDM model.  Among the extensions considered, non-standard effective neutrino species  and non-standard equation of state parameter for dark energy,   bring the CMB and the local measures closer.      
                      
\begin{figure}[!h]
        \centering
\hspace*{-.4cm}                
            \includegraphics[scale=0.3]{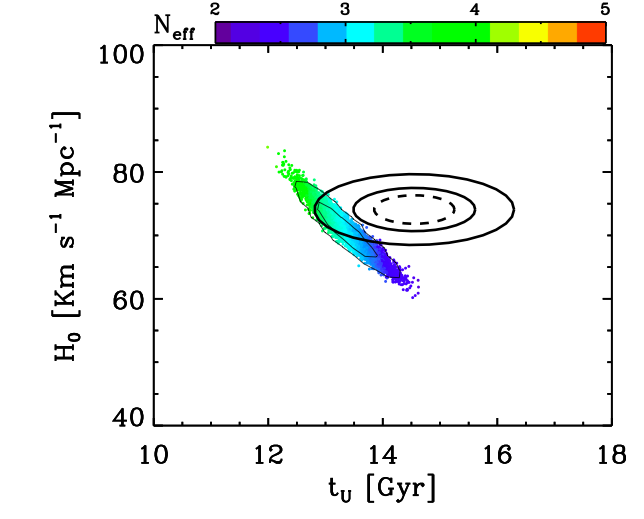}
                \includegraphics[scale=0.3]{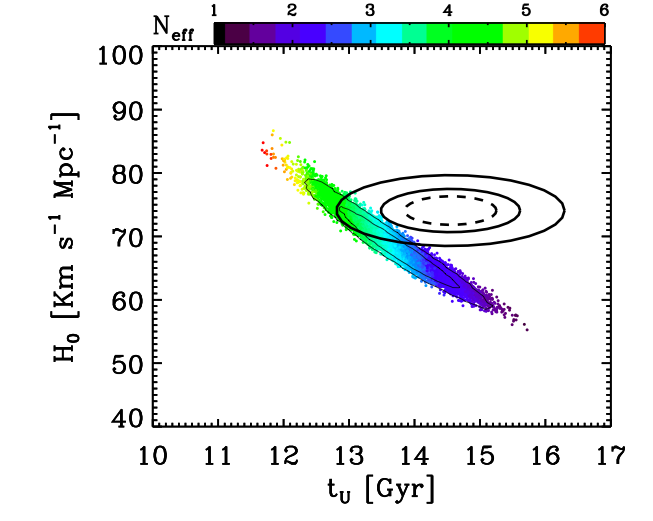}
                  \includegraphics[scale=0.3]{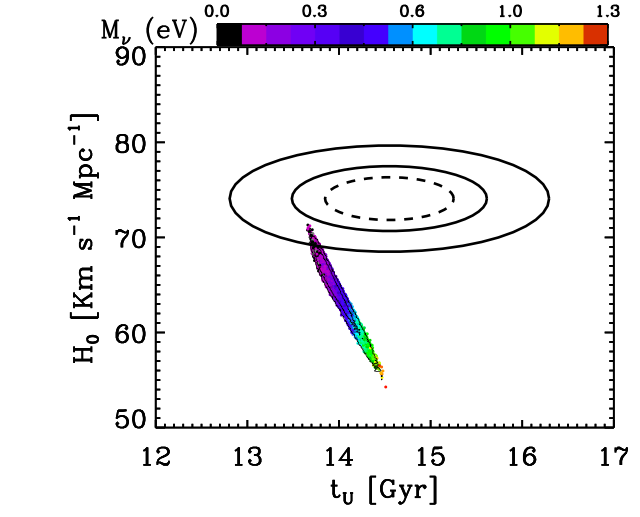}
                \includegraphics[scale=0.3]{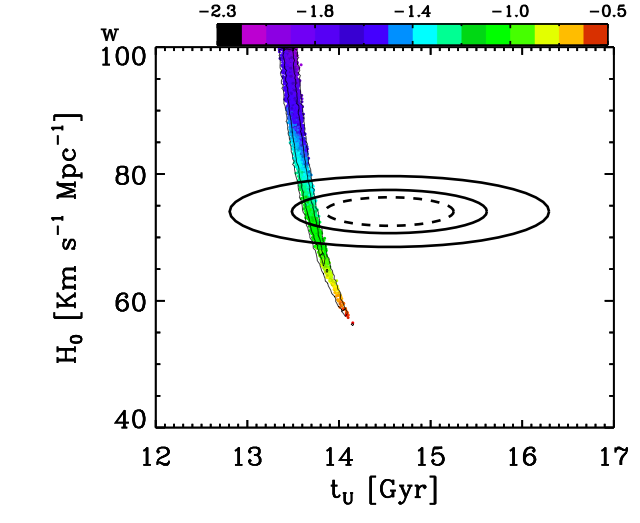}
               \caption{\label{fig:scatterplot} Posterior distributions in the $t_U$--$H_0$ plane for local and CMB measurements. A random sub-sample of the  CMB MCMC points has been shown color-coded by the value of the effective number of neutrino species $N_{\rm eff}$  on the top panels. In the left panel the primordial helium fraction is kept fixed at the nucleosynthesis value  while on the right it is left as a parameter which is then marginalized. A larger $N_{\rm eff}$ value brings in better agreement the $H_0$ determinations (but the agreement worsens for $t_U$). On the bottom panels we show  the $M_{\nu}$ (left) and $w$ (right) extension to the $\Lambda$CDM model.}
 \end{figure}
 
This is further illustrated in Fig.~\ref{fig:scatterplot} (top panels) for the effective number of neutrino species ($N_{\rm eff}$) extension to  $\Lambda$CDM. A  $N_{\rm eff}$ value larger than the standard $3.046$ brings in better agreement the $H_0$ determinations, but the agreement worsens for $t_U$.

Similarly the bottom left  panel of  Fig.~\ref{fig:scatterplot} shows the effect for  the (total) neutrino mass, $M_{\nu}$,  extension and the bottom right panel for  the  non-standard equation of state parameter, $w$, extension of $\Lambda$CDM.  Clearly the constraints on $M_{\nu}$ obtained using the local  $H_0$ determination are very tight thanks to the (local) $H_0$ central value.  Also values of $w<-1$ bring the two $H_0$ estimates in better agreement, in this case, the $t_U$ constraint is not useful given the direction of the CMB degeneracy.

\subsection{Is there Evidence for tension?}
\label{sec:tension}
We can use the method outlined in Sec.~\ref{sec:evidencetension} to quantify if there is Evidence  for tension  between the local and CMB measurements in the $H_0$--$t_U$ plane.
For the ``base" $\Lambda$CDM model  we obtain $\mathcal{T}= 53$, $\ln\mathcal{T}=3.95$, i.e. that the null hypothesis, $H_1$, is disfavoured with odds of roughly 1 to 50. This indicates  a  {\it strong} evidence for tension. We estimate that the numerical error on $\ln\mathcal{T}$, due to the fact that the CMB posterior is not known with infinite precision but sampled by MCMC, is $0.1$.

To gain a physical intuition about this result, let us assume that the {\it Planck's} one-dimensional posterior distribution for $H_0$  (marginalized over all other parameters) and that for  $t_U$  are Gaussian. This is a  good approximation for the $\Lambda$CDM model.
The local age determination is in good agreement with {\it Planck's}: $\ln\mathcal{T}=0.34$. However for   $H_0$ we obtain  $\ln\mathcal{T}=3.5   $\footnote{$\ln\mathcal{T}=3.5$ is obtained also using the actual Planck posterior as sampled by the  MCMC}. Thus the tension between CMB and local measurements is entirely due to $H_0$.

This  {\it strong} evidence for tension is a signal that  caution must be exercised  if  the results of the two experiments are to be  combined, and the result of this combination should be interpreted with care. We will proceed examining the two cases {\it a)} and {\it b)} outlined above in turn.

It is important to note that this measurement of tension is model-dependent and the value reported here  applies only to the $\Lambda$CDM model.
Even in simple extensions of the models, $\mathcal{T}$ can differ widely. We will return to this in Sec. \ref{sec:extending}. 

\subsection{Option ``Blame the measurement": Interpretation in terms of unknown errors}

Discarding measurements  (or combining measurements)  can be seen as two special cases of the use of hyper-parameters, as described in \cite{Bridle}.  The Planck team noticed some tension between Planck data and the $H_0$ measurement and argued that the local measurement is more likely  to be affected by some unknown systematic than the CMB. For this reason they discard $H_0$ when combining Planck with other measurements.  They nevertheless also provided results and MCMC outputs   for data  combinations that include $H_0$.
There could however be a spectra of intermediate possibilities where the $H_0$ measurement is combined but downweighted by a factor $\alpha$ (or equivalently its error increased by $1/\alpha$). Clearly $\alpha=1$ corresponds to doing  the standard joint analysis and $\alpha=0$ to excluding the measurement.

\begin{figure}[t]
        \centering
                \includegraphics[scale=0.38]{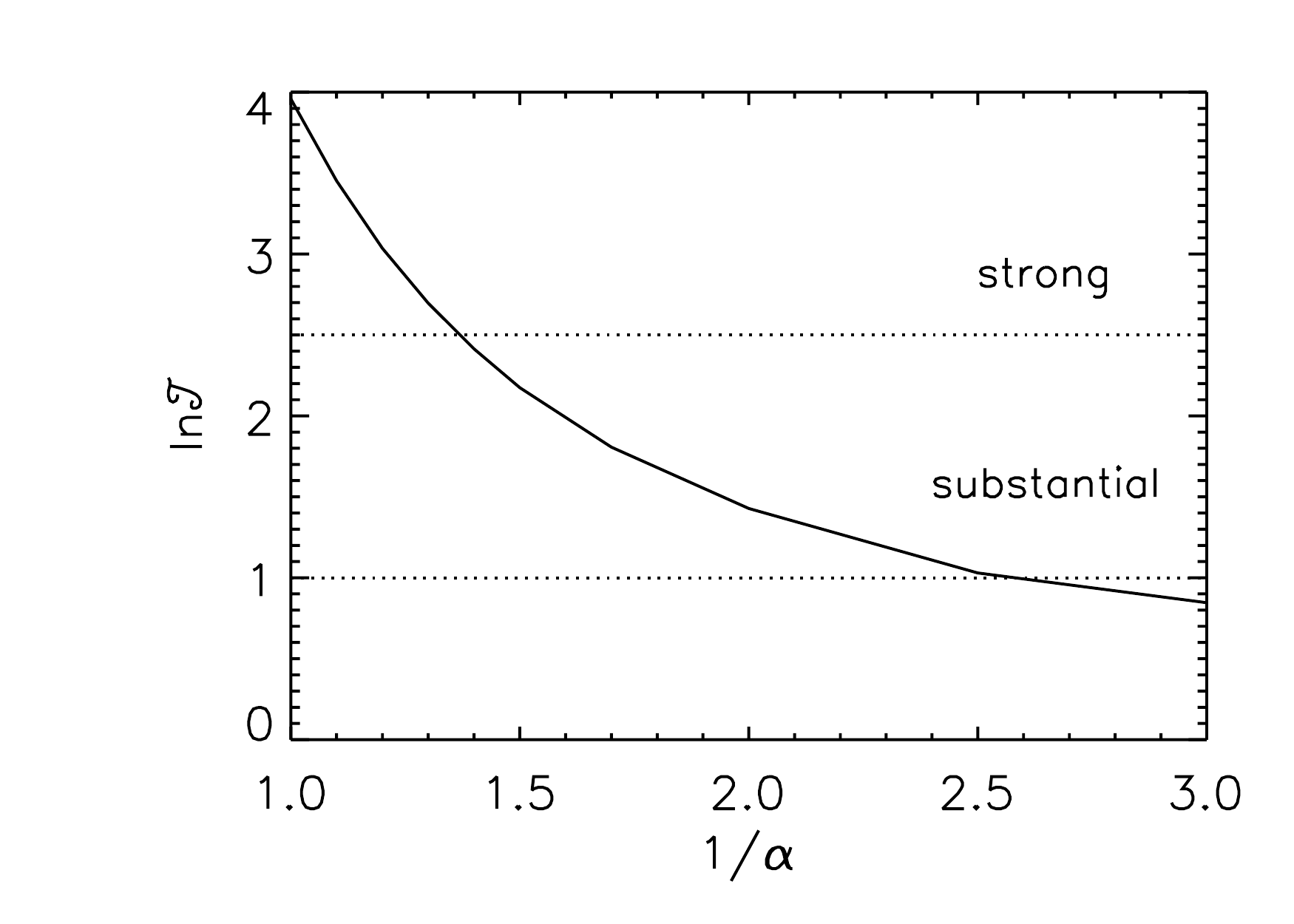}   
                 \includegraphics[scale=0.38]{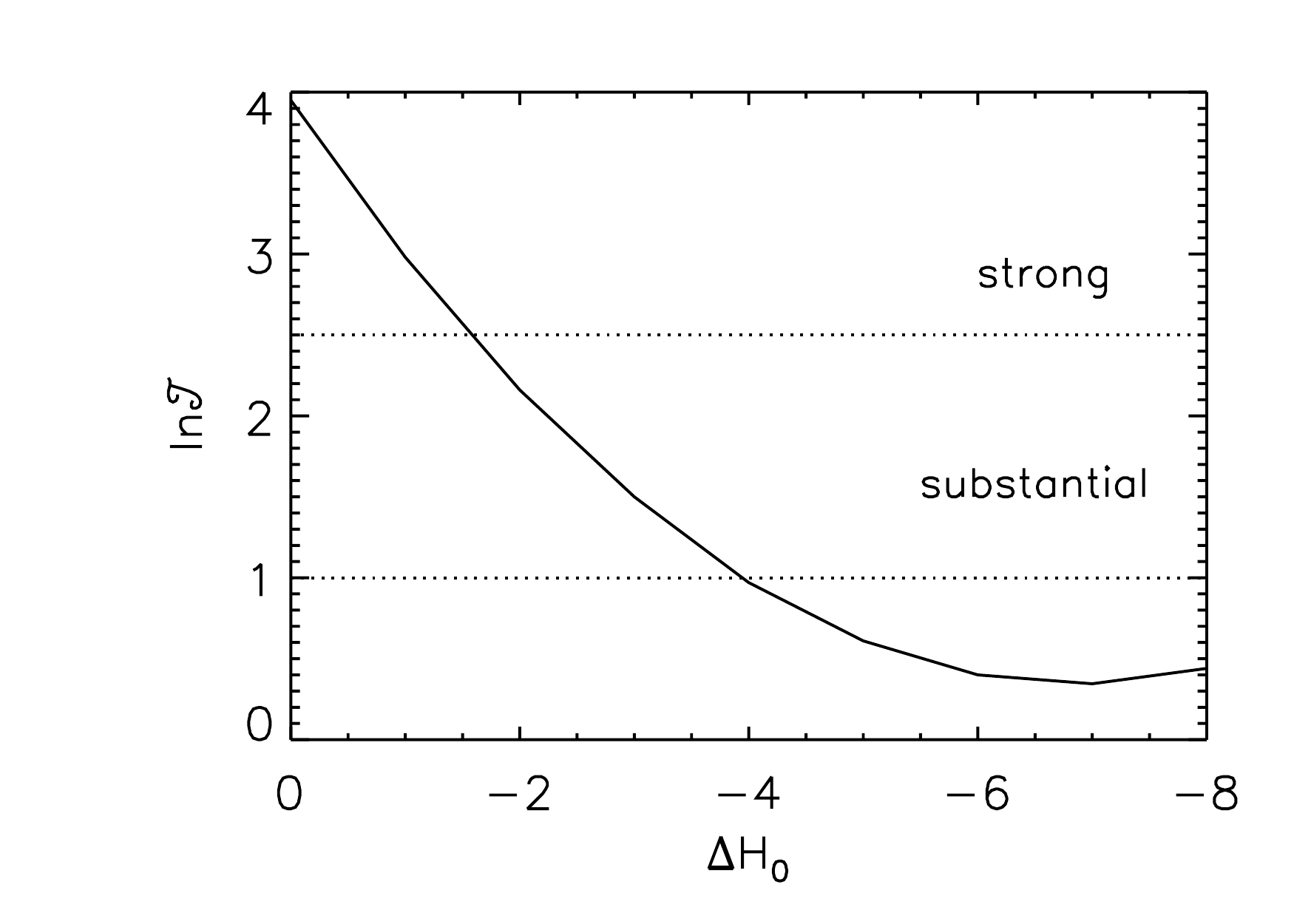}           
               \caption{\label{fig:alpha}  $\ln\mathcal{T}$ for Planck and the local measurements as a function of the factor downweighting the measurement of $H_0$ or as a function of a shift in $H_0$. }
 \end{figure}

Here we follow this train of thoughts and increase the $H_0$ errors to find out what correction would be needed to reduce the tension $\ln \mathcal{T}$ in the $H_0-t_U$ plane, to a more ``comfortable'' level. This is shown in Fig. \ref{fig:alpha}.
Of course we could have decided to downweight the  CMB and/or the age measurements.  The age measurements and the CMB are in good agreement  with each other--but $t_U$ has still large error-bars--, so we would have had to downweight both. We decided here to follow the reasoning of the Planck team, downweight $H_0$ and interpret the consequences.
Another possibility is to model an --unknown-- systematic error by shifting the $H_0$ measurement by $\Delta H_0$ and see what shift is needed to significantly reduce the tension. This is also shown in Fig. \ref{fig:alpha}.
 
This figure shows that an increase in the $H_0$ error by 30-40\% would be needed to reduce  $\ln\mathcal{T}$ to {\it substantial} but a factor 2.5 would be needed for {\it not significant}. On the other hand a shift of the central value by little less  than 1$\sigma$ would bring $\ln\mathcal{T}$ to {\it substantial} but a shift of  almost $\sim 2 \sigma$ would be needed for it to become {\it not significant}.
 
\subsection{Option ``blame the model": Extending the ``base" model}
\label{sec:extending}
The values of the tension obtained so far are valid within the $\Lambda$CDM model.  There may be simple extensions of this model that reduce or remove tension.

\subsubsection{Does tension  decrease in $\Lambda$CDM extensions?}
\label{sec:tensionext}
In the first line of table \ref{tab:modelselection} we report $\ln\mathcal{T}$ for simple extensions of the $\Lambda$CDM model.  We see that including the curvature, $\Omega_K$ as a parameter or including a  parameter for non-zero  neutrino mass, $M_{\nu}$, increases $\ln\mathcal{T}$, thus disfavoring these model extensions. The two model extensions, however, should not be considered on the same footing: inflation strongly motivates the assumption that the Universe is flat (and therefore there is no need for the extra curvature parameter). On the other hand neutrino oscillations indicate that neutrinos have a non-zero mass, $M_{\nu}$ must be larger than about $0.05$ eV  and the cosmologically-independent upper limit is $\sim 2$eV. Therefore  assuming a $\Lambda$CDM model with  massless neutrinos, or with $M_{\nu}$ fixed at a value close to its lower limit, is not really motivated and one could argue that  $\Lambda$CDM$+M_{\nu}$ model should be the ``base'' model.

The addition of the effective number of neutrino species $N_{\rm eff}$ as a parameter brings the tension down to {\it substantial}, and the addition of the equation of state for dark energy parameter $w$ brings it to {\it not significant}.  This indicates that these model extensions are  particularly interesting and warrant more investigation (under case {\it b)}, of course).  Contrary to the  $\Lambda$CDM$+M_{\nu}$ case,  these model  extensions do not have other strong experimental motivations. 

It is interesting to compare these findings with the standard  Bayesian model selection. 

\subsubsection{Bayesian model selection}
We carry out the standard Bayesian model selection computing the Bayes factor (i.e. the Evidence ratio) between the ``base" $\Lambda$CDM model and its extensions.  We start by producing versions of  the  relevant MCMC chains  importance-sampled  with the local constraints.
We  then follow Ref. \cite{Plancknu}  to compute the Evidence ratio between two nested models from an MCMC output via the Savage-Dickey density ratio. The Evidence ratio for  $\Lambda$CDM extensions involving neutrino properties differ from those in Ref. \cite{Plancknu} because here we also include the $t_U$ determination. The results are reported in Table \ref{tab:modelselection} (second line). We see that  even combining {\it Planck} with local universe measurements, there is never substantial or  strong evidence for the model extension and in some  cases the simpler model $\Lambda$CDM is strongly favored. 

\begin{table}[tb]
\centering
 \begin{tabular}{|l|ccccc|}
 \hline
model extension& $w$&$\Omega_k$&$N_{\rm eff}$&$M_{\nu}$&$N_{\rm eff} +Y_P$ \\
\hline
  $\ln \mathcal{T}$ &  0.74 &5.24 & 1.94& 4.5 & 2.2 \\
$\ln E_{\Lambda CDM}/E_{\rm extension}$&-0.72&3.70&-0.27(P)&3.45&1.93(P)\\

  \hline
\end{tabular}
 \caption{Tension ($\ln \mathcal{T}$) between Planck and  the local measurements in $\Lambda$CDM extensions and Evidence ratio between the $\Lambda$CDM model and its extensions.  For the tension, numbers above 2.5 signify {\it strong} tension and above 5 {\it highly significant} tension. For the Evidence ratio positive numbers mean that the simpler model is preferred, negative numbers that the more-complicated model is preferred; $|\Delta \ln E|$  should be at least 2.5 to be {\it strong}. The errors on the evidence ratios are in most cases about $\pm 0.02$, reaching $\pm 0.1$ for thinned, post-processed chains (indicated by P). }
 \label{tab:modelselection}
\end{table}

This is not necessarily in contradiction with the  findings of Sec. \ref{sec:tensionext} as the two quantities measure different things and address different questions. In particular  the Bayesian Evidence  is not concerned with whether  at least one of the two models considered  provides a good fit to all the data simultaneously. But this is exactly what  $\ln \mathcal{T}$ does.

Qualitatively however the trends of the two statistics are similar:  the models more disfavored by the Bayesian evidence are those for which the tension increases with respect to the $\Lambda$CDM case.  The models that decrease the tension most are those that favor  (only ever so slightly) the more complex model.

\begin{figure}
        \centering
                \hspace*{-2cm}
                \includegraphics[width=1.3\columnwidth]{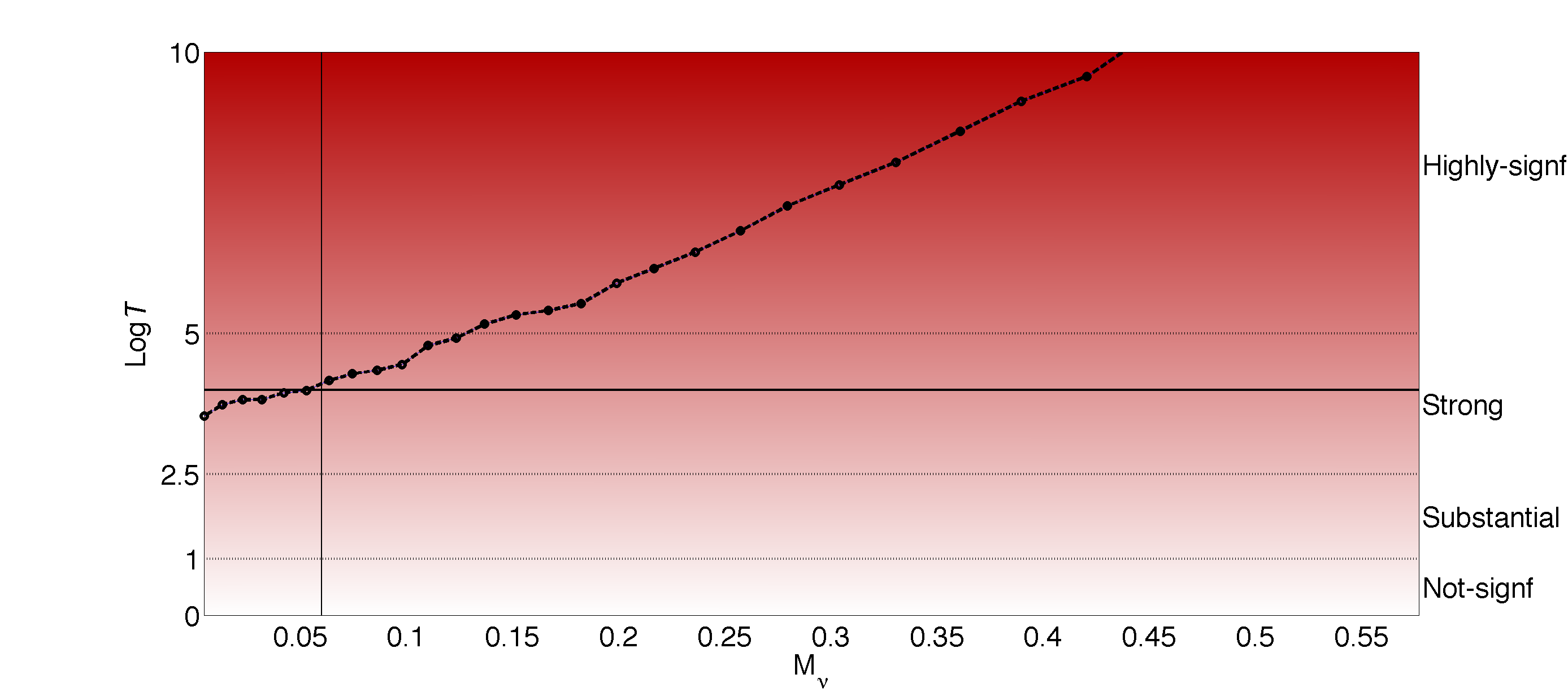}
                \hspace*{-0.8cm}
                \includegraphics[width=1.22\columnwidth]{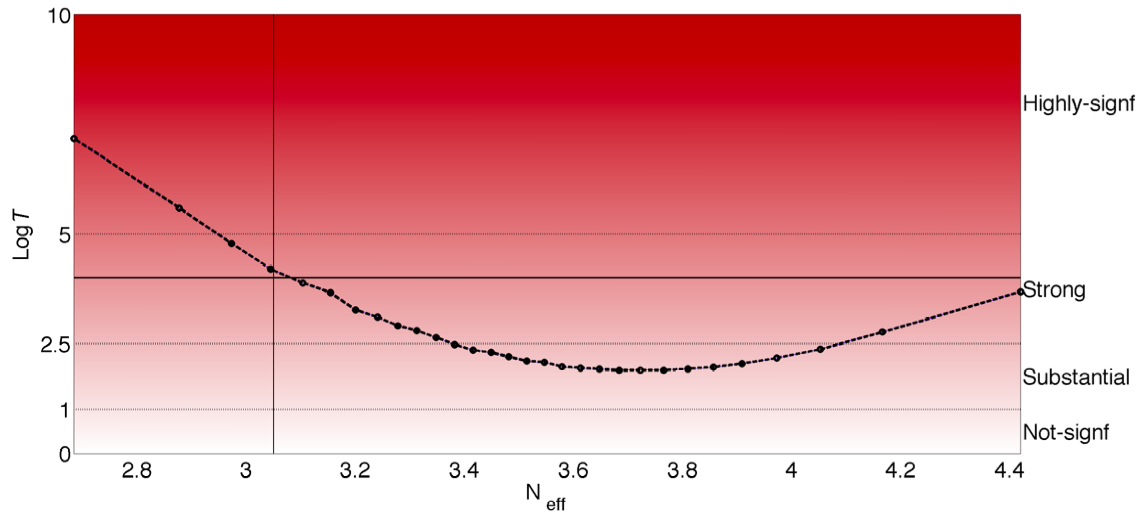}             
               \caption{\label{fig:nuparam} $\ln\mathcal{T}$  as a function of $M_{\nu}$ (top panel) and as a function of $N_{\rm eff}$ (bottom panel).  The black solid line corresponds to the $\Lambda$CDM value, odds $\sim1\,:\,50$. Note that for values of $M_{\nu}$ higher than $0.15$ eV, the tension between local measurements and Planck derived values increases to {\it highly significant} (odds $\sim 1:150$). This  indicates that the degenerate  hierarchy for the neutrino mass spectrum is highly disfavored and that normal hierarchy is preferred over the inverted one.  However no value of $M_{\nu}$ yields {\it non-significant'}  or even {\it substantial} tension}
 \end{figure}

\begin{figure}
        \centering
        \hspace*{-2cm}
                 \includegraphics[width=1.33\columnwidth]{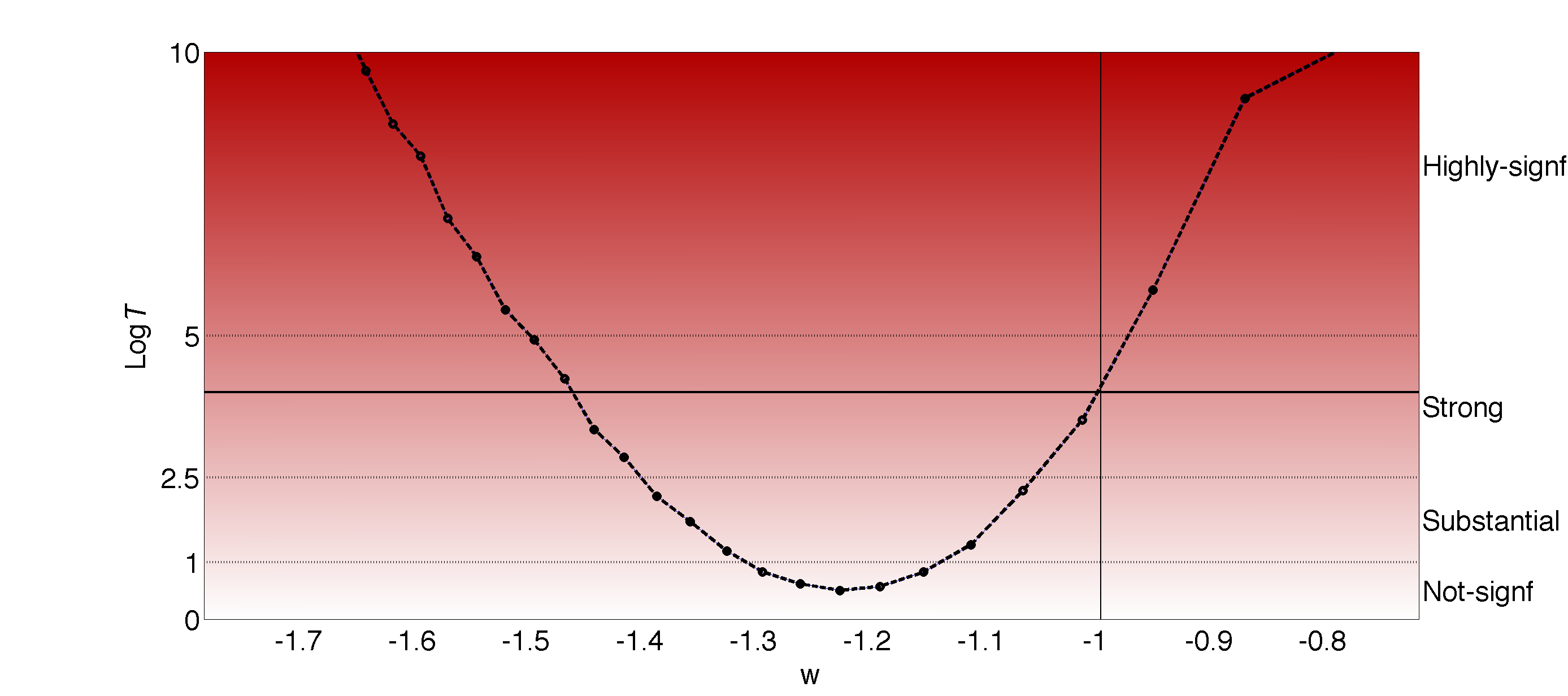}
                 \hspace*{-1.35cm} 
                 \includegraphics[width=1.33\columnwidth]{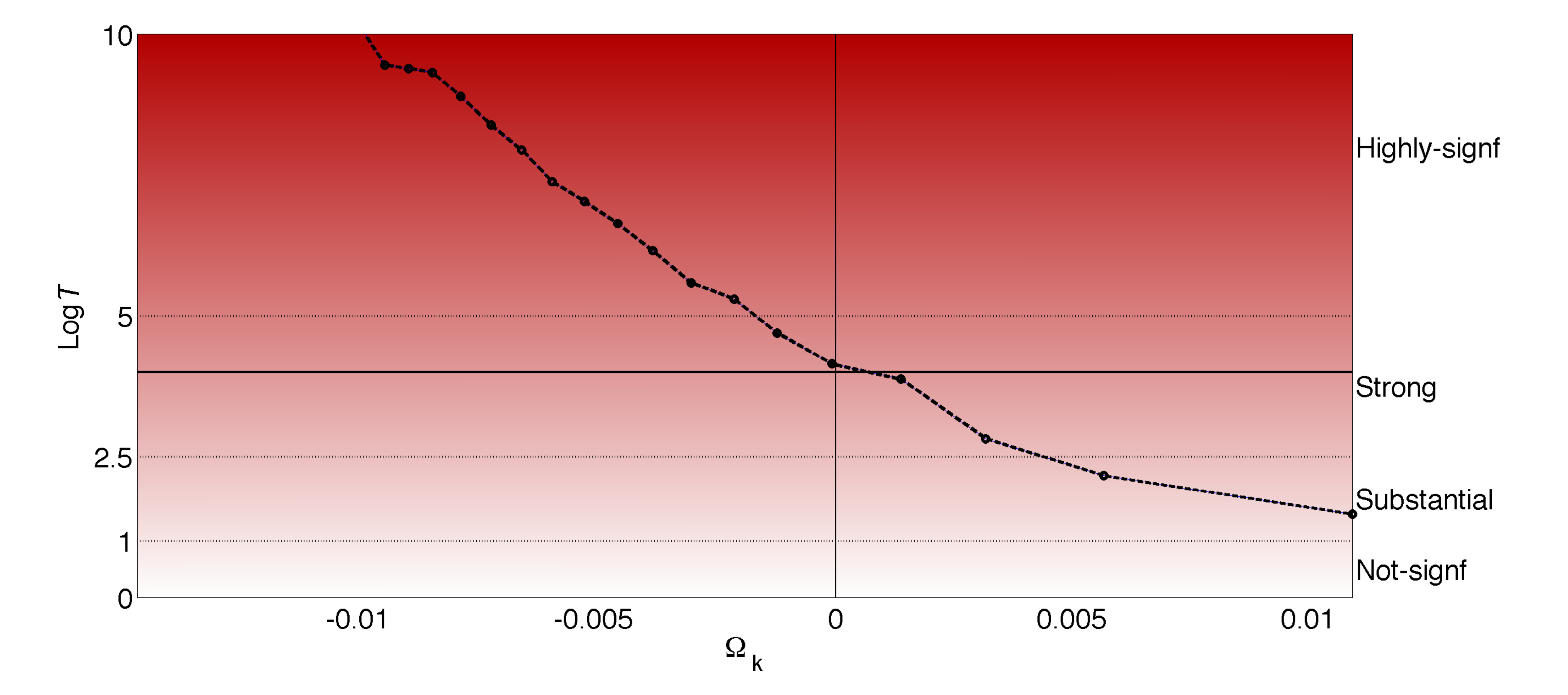}                 
               \caption{\label{fig:DEparam} Same as Fig.~\ref{fig:nuparam} but now changing the equation of state parameter for dark energy $w$ (top panel) and the curvature parameter $\Omega_k$ (bottom panel).  A slightly ``phantom"  ($w<-1$) value for the equation of state parameter bring the tension to more comfortable ``substantial" and  and values around $w=-1.2$ to ``not significant".  Alternatively a slightly positive curvature $\Omega_k>8\times 10^{-3}$ brings the tension to  ``not significant".}
 \end{figure}

\subsubsection{Tension and its implications for models and parameters}
The fact that for some $\Lambda$CDM extensions tension is reduced (or even eliminated) but not for others, suggest that we can use the quantity $\ln \mathcal{T}$ for  deriving  constraints on  the  parameters of the model extension.

For example we can ask: ``if we had a compelling reason to fix the extra parameter  to a given value, what would the tension be?" Clearly, parameters  values that yield ``very strong" tension  should be disfavored.
 This is what we show in Figs. \ref{fig:nuparam}-\ref{fig:DEparam}.
 
 In Fig. ~\ref{fig:nuparam} we show results for $\Lambda$CDM model extensions involving non-standard neutrino properties. 
We find that the lower the neutrino mass the lower $\ln \mathcal{T}$.  For values of $M_{\nu}$ higher than $0.15$ eV, the tension between local measurements and Planck derived values increases to {\it highly significant} (odds $\sim 1:150$). This indicates that the degenerate  hierarchy for the neutrino mass spectrum is highly disfavored and that normal hierarchy is preferred over the inverted one (although no value for $M_{\nu}$ yield {\it not significant}  or even {\it substantial} tension). 
 
Values of $N_{\rm eff}$ larger than the standard value ($3.4<N_{\rm eff}<4.1$) reduce $\ln \mathcal{T}$,  but values $N_{\rm eff}>4.6$ make it {\it highly significant}. Note that for no range of $N_{\rm eff}$ values  the tension is  {\it not significant}. This is because among  all the $N_{\rm eff}$ values that are a good fit to Planck data,  values higher than the fiducial improve the  fit to   $H_0$ data  but worsen the fit to $t_U$.
 
 In Fig.~\ref{fig:DEparam}  we show results for  the dark energy equation of state parameter $w$, and curvature $\Omega_k$, $\Lambda$CDM model extensions. Both these model  extensions have parameter values that make the tension {\it not significant}.
 A slightly ``phantom"  ($w<-1$) value for the equation of state parameter brings the tension to more comfortable {\it substantial} and   values around $w=1.2$ to {\it not significant}.  Alternatively a slightly positive curvature $5\times 10^{-3}<\Omega_k<1.5\times 10^{-2}$ brings the tension to  a more comfortable  ({\it substantial}) level.
 
\section{Discussion and Conclusions}
\label{sec:conclusions}

We have analised cosmology independent, local measurements of $H_0$ and $t_U$ and the Planck derived values  for these quanities within the $\Lambda$CDM model and in simple extensions of it. We started by introducing and developing  a statistic,  the tension $\mathcal{T}$, to determine when two posterior distributions of  parameters are in tension and  to quantify the level of tension.  
We argue that the tension is based on the Bayesian evidence and the Bayes factor and, as such, it can be interpreted using the popular Jeffreys' scale. To give an intuition about the meaning of $\ln \mathcal{T}$ values we should bear in mind that if we have two 1-dimensional Gaussian  distributions with unit variance, a shift of the central value of one of the two distributions  of $2 \sigma$ would give $\ln \mathcal{T}=1$ which is the threshold between {\it not significant} and {\it substantial}; a shift of more than  $3 \sigma$ would yield {\it strong} tension and  of more than $4.5 \sigma$ to give {\it highly significant} tension.

We  then have shown that, in agreement with the(recently)  commonly accepted wisdom, in the framework of the $\Lambda$CDM model, these two determinations are in {\it strong} tension (odds $\sim 1\,:\,50$). In our naive Gaussian interpretation above these odds corresponds to a shift of one of the two unit-variance Gaussians if $\sim 4 \sigma$. 

We recognize that two broad classes of explanations for this  result are possible. ``Blame the measurement": one (or more) of the measurements have errors that are unaccounted for or "blame the model":  extensions of  base $\Lambda$CDM model  should be considered.   

Following the interpretation suggested by the {\it Planck} team we have then explored how to alleviate the tension by exploring by how much the errors and/or the central value of $H_0$ needs to be changed in order to  alleviate the tension to tolerable levels, odds $> 1:3$ ($>1:10$). We found that this can be achieved if the error bars have been underestimated by a factor $2.5$ (30-40\%) or the central value is wrong by $4$ ($\sim 2$) km s$^{-1}$ Mpc$^{-1}$. 
Having quantified this, we leave it to the experts in  the field of the $H_0$ measurements to judge whether  this is a realistic possibility, keeping in mind that two independent measurements, in excellent agreement with each other, were suitably combined to obtain the adopted $H_0$ constraint.

We have then explored extensions to the $\Lambda$CDM that can alleviate this tension. We found that several extra parameters can achieve this: allowing for phantom values of the equation of state of dark energy ($w \sim -1.2$) or allowing a small positive curvature ($7\times 10^{-3}<\Omega_k<1.5\times 10^{-2}$).

No values of the effective number of species reduces the tension to not significant, but values around $3.6$--$3.8$ get close to that. 
An interesting finding is that when using the neutrino mass as an additional parameter for the $\Lambda$CDM model,  although no values for the total neutrino mass reduce the tension to {\it not significant} (odds $1\,:\,3$), a total  mass above  $0.15$ eV makes the tension {\it highly significant} (odds $\sim 1:150$). A consequence of accepting this interpretation of the tension is that  the degenerate neutrino hierarchy is highly disfavoured by cosmological data and the direct hierarchy  is slightly favored over the inverse one.  Of course, if we accept this interpretation of  the ``blame the model" option, this could be the first indication from cosmology for a neutrino hierarchy (e.g., \cite{Jimnu} and refs. therein).

There is one possible explanation for the tension that lies in-between ``blame the data" and ``blame the model" options, that is that we are in a local underdensity. Ref. \cite{Marra} shows that measurements of the local Hubble constant are subject to a cosmic variance error with can be as high as  2\%.  If, simply by cosmic variance,  we happen to live inside an 
underdensity, the local values of $H_0$ could be higher than the cosmological value by this amount. As we have seen in Fig.~\ref{fig:alpha},  such shift would certainly reduce the tension but not remove it. The larger shift needed for this, the authors of \cite{Marra} argue,  would however require a very rare  (i.e., unlikely) fluctuation for the model.

Tantalizingly, observations of the luminosity density as function of redshift \cite{ukids} suggest that  we might be located in a $\sim 300$ Mpc/h local under density. It remains to be  assessed wether such under density is due to cosmic variance, can be accommodated in the $\Lambda$CDM scenario and thus only adds a (small) systematic correction to the local $H_0$  measurement (case ``blame the measurement"). On the other hand, if  the underdensity is large enough,  it could  eliminate the tension but  at the expense of requiring  indeed an  (interesting) extension to the $\Lambda$CDM model (case ``blame the model").

We have focussed our analysis only on local, cosmology independent, measurements and the CMB derived local universe within a given model. We have chosen not to use other cosmological probes as BAO or Supernova as they are not local measurements and  not as mature from a theoretical point of view (and not as powerful when used alone) as the CMB  to be used as high-redshift measurements.
Moreover we have not used the high-$\ell$ CMB data ({\it highL}) nor the CMB lensing information because of possible internal tension with the {\it Planck} temperature data (although we have checked that adding {\it highL} data does not change significantly or qualitatively our findings).  The {\it Planck} experiment statistical power will improve drastically when polarization data will be included in the analysis over the next year. The inclusion of  these extra {\it Planck} data will cement the CMB constraints (strengthening or weakening the above findings).  

After that, if tension remains, a way forward  is to improve local data to the \% level. The ages of the oldest stars show no tension with Planck CMB data, but error-bars are still relatively large; thus much more accurate ages from the GAIA mission will help elucidate  the situation. \\

{\bf Acknowledgments:} LV   is supported by European Research Council under the European Community's Seventh Framework Programme grant FP7-IDEAS-Phys.LSS. LV and RJ acknowledge Mineco grant FPA2011-29678- C02-02.
We thank Hiranya Peiris and Arthur Kosowsky for discussions and comments on an early version of this work. 
We acknowledge the use of the Legacy Archive for Microwave Background Data Analysis (LAMBDA), part of the High Energy Astrophysics Science Archive Center (HEASARC). HEASARC/LAMBDA is a service of the Astrophysics Science Division at the NASA Goddard Space Flight Center. This work is based on observations obtained with Planck (http://www.esa.int/ Planck), an ESA science mission with instruments and contributions directly funded by ESA Member States, NASA, and Canada. The development of Planck has been supported by: ESA; CNES and CNRS/INSU-IN2P3-INP (France); ASI, CNR, and INAF (Italy); NASA and DoE (USA); STFC and UKSA (UK); CSIC, MICINN and JA (Spain); Tekes, AoF and CSC (Finland); DLR and MPG (Germany); CSA (Canada); DTU Space (Denmark); SER/SSO (Switzerland); RCN (Norway); SFI (Ireland); FCT/MCTES (Portugal); and PRACE (EU). We acknowledge the use of the Planck  Legacy Archive.

\appendix
\section{How many bits of information has Planck added to WMAP?}
\label{sec:kld}
The  Kullback-Leibler divergence \cite{KL} $D_{KL}({\cal P} || {\cal W})$, as mentioned in Sec. \ref{sec:methods}, quantifies the information content that one distribution (${\cal P}$) adds to the other one (${\cal W}$);  in other words,  how many bits of information  are needed if we are given the second distribution (${\cal W}$) and we  want to recover the first one (${\cal P}$), or,  how many bits are lost if one were to  use ${\cal W}$ to approximate ${\cal P}$. Here we are interested in using this statistic to  quantify how much new information has Planck added  to WMAP. We will use $D_{KL}({\cal P} || {\cal W})$, where ${\cal P}$, ${\cal W}$ denote the {\it Planck}(+WP)  and WMAP posterior distributions,
\begin{equation}
  D_{KL}({\cal P} || {\cal W})=\int_x\log_2\left(\frac{{\cal P}(x)}{{\cal W}(x)}\right){\cal P}(x)dx\,.
\end{equation}
The logarithm in base $2$ means that the information is expressed in bits.
The above equation can be interpreted and applied in several ways for example it can be applied to the  1 dimensional distribution for one parameter marginalized over all the other ones, or as the joint distribution of two parameters or as the joint, multivariate distribution of all the cosmological parameters.
Here we present only few selected cases, see table \ref{tab:KL}. In the  first block  we report the $D_{KL}$ parameter by parameter in the ``base" $\Lambda$CDM model. The 1 dimensional  posterior used has been marginalized over all other parameters. The second block includes  the $D_{KL}$ for simple extensions of the $\Lambda$CDM model. We report the  one dimensional  posterior  of the extra parameter marginalized over all other parameters. First we report the results for {\it Planck}+ WP, then the addition of {\it highL} and the further addition of {\it lensing}.

These numbers should be interpreted as follows: they represent the extra bits of information added, which do not need to be whole numbers, by {\it Planck} over WMAP in bits in base $2$. So, for example, if $1$ is added, the information is multiplied by a factor $2^1$. In the case of $N_{\rm eff}$, Planck has increased the information of WMAP by a factor $2^{1.7} = 3.2$. This is the parameter for which most information has been added, while for the other parameters Planck has increased information by about a factor $2$.  
If we include Planck's {\it highL},  the one-parameter $\Lambda$CDM numbers are mostly unchanged except  for  the (scalar) power spectrum spectral slope, $n_s$, which
 becomes $1.36$. Including also {\it lensing} does not add significant information for this model. This is not the case for the $\Lambda$CDM model extensions. The effect of the previously mentioned ``tension" between {\it Planck +WP+highL} and {\it lensing} can be seen in the $M_{\nu}$ column.

Table \ref{tab:KL} should not be interpreted  as that Planck has improved over WMAP by a factor 2. In fact the reported numbers are for individual parameters, but  the experiment measures all the parameters of the model (typically 6 or 7) as well as non-cosmological but nevertheless  astrophysically interesting  parameters, and all are improved. 

\begin{table}[tb]
\centering
 \begin{tabular}{lccccc}
 \hline
 \hline
 $\Lambda$CDM& $w_b$&$w_c$&$n_s$&$H_0$& $t_U$\\
 \hline
1D for parameter &1.35&1.63&1.09&1.21&0.83\\
 \hline
 &&&&&\\
 \hline
$\Lambda$CDM Extension& $\Omega_k$&$w$&$N_{\rm eff}$&$M_{\nu}$&$$\\
 \hline
 1D for parameter&0.67&1.05&1.70&0.39& \\
 +{\it highL} &0.82&1.18&1.90&0.91&\\
 +{\it lensing} &1.07&1.15&1.94&0.38&\\
 \hline
\hline
\end{tabular}
 \caption{Kullback-Leibler divergence of {\it Planck} from WMAP, ${\cal D}_{KL}({\cal P}||{\cal W})$. Here $w_b$ denotes the physical density of baryons, $w_c$ the physical density of cold dark matter, $n_s$ the primordial matter power spectrum spectral slope, $H_0$ the Hubble  constant, $t_U$ the age of the Universe, $\Omega_k$ the curvature parameter, $w$ the dark energy equation of state parameter, $N_{\rm eff}$  the effective number of neutrino species and $M_{\nu}$ the total neutrino mass. The ``base" data set  corresponds to {\it Planck +WP}; the addition of {\it highL} and {\it lensing} are also considered.}
 \label{tab:KL}
\end{table}

\bibliographystyle{model1-num-names}

\begin{thebibliography}{00}
 
 \bibitem[Verde et al. (2013)]{local} Verde L., Jimenez R., Feeney, S., Phys. Dark Universe, 2013.
 
 \bibitem[Cox (1946)]{Cox}  R. T. Cox, 
 Am.J.Th.Phys 14 (1946) 1.
 
 \bibitem[Jaffe (1996)]{Jaffe96}
  A.~H.~Jaffe,
  Astrophys.\ J.\  {\bf 471} (1996) 24
  [astro-ph/9501070].
 
 
 
\bibitem[Hobson et al.(2002)]{Hobson} Hobson, M.~P., Bridle, 
S.~L., \& Lahav, O.\ 2002, MNRAS, 335, 377 

\bibitem[Liddle (2002)]{Liddle} 
  A.~R.~Liddle,
  Mon.\ Not.\ Roy.\ Astron.\ Soc.\  {\bf 351}, L49 (2004)
  [astro-ph/0401198].
 
 
\bibitem[Parkinson et al. (2006)]{Parkinson} 
  D.~Parkinson, P.~Mukherjee and A.~R~Liddle,
  Phys.\ Rev.\ D {\bf 73}, 123523 (2006)
  [astro-ph/0605003].
  
\bibitem[Trotta \& Melchiorri (2004)]{Trotta1} 
  R.~Trotta and A.~Melchiorri,
  Phys.\ Rev.\ Lett.\  {\bf 95}, 011305 (2005)
  [astro-ph/0412066].
  
  \bibitem[Trotta (2007)]{Trotta2}
  R.~Trotta,
  Mon.\ Not.\ Roy.\ Astron.\ Soc.\  {\bf 378} (2007) 72
  [astro-ph/0504022].
  
 \bibitem[Planck Collaboration (2013a)]{Planckbasicresults}
  P.~A.~R.~Ade {\it et al.} [Planck collaboration],
  arXiv:1303.5062 [astro-ph.CO].
 
 \bibitem[Planck Collaboration (2013b)]{Planckparameterspaper} P.~A.~R.~Ade {\it et al.}[Planck collaboration],
  arXiv:1303.5076 [astro-ph.CO].
 
 \bibitem[Feeney et al. (2013)]{Feeney2} Feeney S., Peiris H.V., Verde, L., in preparation (2013)
 
 \bibitem[Planck Collaboration (2013c)]{Plancklensingpaper} P.~A.~R.~Ade {\it et al.}[Planck collaboration],
  arXiv:1303.5077 [astro-ph.CO].
  
  
 \bibitem[Riess et al.(2011)]{Riess/etal:2011} Riess, A.~G., Macri, L., 
Casertano, S., et al.\ 2011, ApJ, 730, 119 

\bibitem[Freedman et al.(2012)]{Freedman/etal:2012} Freedman, W.~L., 
Madore, B.~F., Scowcroft, V., et al.\ 2012, ApJ, 758, 24 

\bibitem[Bond et al.(2013)]{Bond} Bond, H.~E., Nelan, E.~P., VandenBerg, D.~A., Schaefer, G.~H., \& Harmer, D.\ 2013, ApJL, 765, L12 

\bibitem[Krauss 
\& Chaboyer(2003)]{chabkrauss} Krauss, L.~M., \& Chaboyer, B.\ 2003, Science, 299, 65

\bibitem{Jimenez:1996at}
  R.~Jimenez, P.~Thejll, U.~Jorgensen, J.~MacDonald and B.~Pagel,
  Mon.\ Not.\ Roy.\ Astron.\ Soc.\  {\bf 282} (1996) 926
  [astro-ph/9602132].

\bibitem[Gratton et al.(2003)]{Gratton03} Gratton, R.~G., Bragaglia, A., Carretta, E., Clementini, G., Desidera, S., Grundahl, F., Lucatello, S.\ 2003.\ Distances and ages of NGC 6397, NGC 6752 and 47 Tuc.\ Astronomy and Astrophysics 408, 529-543.


\bibitem[Imbriani et al.(2004)]{Imbriani} Imbriani, G.,  et al.  2004.\ The bottleneck of CNO burning and the age of Globular Clusters.\ Astronomy and Astrophysics 420, 625-629. 

\bibitem[Haiman(2011)]{Zoltan} Haiman, Z.\ 2011.\ Cosmology: A 
smoother end to the dark ages.\ Nature 472, 47-48. 

\bibitem[Kullback, Leibler (1951)]{KL}Kullback, S., Leibler, R. A., 1951. On information and sufficiency. Annals of Mathematical Statistics 22, 49�86.

\bibitem[Jeffreys (1973)]{Jeffreys}Jeffreys, H., 1973. Scientific Inference. Cambridge University Press.\

\bibitem[Kass \& Raftery (1995)]{kassraftery}  Kass, R.E. and Raftery, A. E., 1995,  Bayes factors.\ JASA 90, 430, 773-795.


\bibitem[Lin (1991)]{Lin}Lin, J., 1991. Divergence measures based on the Shannon entropy. IEEE Transactions on Information Theory 37, 145�151.


\bibitem[Lahav et al.(2000)]{Bridle} Lahav, O., Bridle, S.~L., 
Hobson, M.~P., Lasenby, A.~N., \& Sodr{\'e}, L.\ 2000, MNRAS, 315, L45 


\bibitem[Verde et al. (2013)]{Plancknu} Verde L., Feeney S., Peiris H. V., Mortlock D., 2013, arXiv 
 
\bibitem[Jimenez et al. (2010)]{Jimnu} R.~Jimenez, T.~Kitching, C.~Pena-Garay and L.~Verde,
  JCAP {\bf 1005} (2010) 035
  
\bibitem[Marra et al. (2013)]{Marra} Marra, V., Amendola L., Sawicki I., Valkenburg, W., 2013, Phys. Rev. Lett. 110, 241305 

\bibitem[Keenan et al. (2013)]{ukids} Keenan, R.C., Barger,A., J., Cowie L.L., 2013, arXiv:1304.2884 
  
\end{thebibliography}

\end{document}